\documentclass[onecollarge, referee]{svjour2}

\usepackage{amsmath} 
\usepackage{amssymb} 
\usepackage{amsfonts}
\usepackage[dvips]{graphicx} 
\usepackage[]{epsfig} 
\begin{document}
\bibliographystyle{spmpsci} 

\title{How interface geometry dictates water's thermodynamic signature in hydrophobic association} 
\author{Joachim Dzubiella} 
\institute{Helmholtz Zentrum Berlin f\"ur Materialien und Energie, Hahn-Meitner-Platz~1, 14109 Berlin, Germany, and Department of Physics, Humboldt University Berlin, Newtonstr.~15, 12489 Berlin, Germany }
\thanks{joachim.dzubiella@helmholtz-berlin.de}

\maketitle

\begin{abstract}

As a common view the hydrophobic association between molecular-scale binding partners is supposed to 
be dominantly driven by entropy.  Recent calorimetric experiments and computer  simulations heavily challenge 
this established paradigm by reporting that water's thermodynamic signature in the binding of
small hydrophobic ligands to similar-sized apolar pockets is enthalpy-driven. Here we show with 
purely geometric considerations that this controversy can be resolved if the antagonistic  effects of concave and convex bending on water interface thermodynamics are properly taken  into account.  A key prediction of this continuum view is that for fully complementary binding of the convex ligand to the concave counterpart, water shows a thermodynamic signature very similar to planar (large-scale) hydrophobic association, that is, enthalpy-dominated, and hardly depends on the particular pocket/ligand geometry. A detailed comparison to recent simulation data qualitatively supports the validity of our perspective down to subnanometer scales. Our findings have important implications for the interpretation of thermodynamic signatures  found in molecular recognition and association processes. Furthermore, traditional implicit solvent models may benefit from our view with respect to their ability to predict binding free energies and entropies.

\end{abstract}

\section{Introduction}

One of the major driving forces in biomolecular association and self-assembly is the hydrophobic 
effect, mediated by structural rearrangements of the surrounding water molecules and thus 
ubiquitous in every biological reaction.  Much progress has been made in the last decade in understanding
the hydrophobic effect, in particular, by the distinction between the qualitatively different hydration 
behavior at small and large length scales~\cite{Lum:1999,Chandler:2005}.  Water's bulk-like 
hydrogen network is only moderately distorted around small, {\it convex} solutes upon hydration while strongly 
penalized by configurational freedom.   Consequently, below a cross-over scale of $\sim$1 nm water's thermodynamic signature to hydrophobic  association is entropy dominated, in contrast to large-scale enthalpy-driven association. This is 
often exemplified by the hydration or association of small, apolar solutes, such as methane. It has thus become a 
common view that small-scale hydrophobic association is entropy-driven. 

That the reasoning above may not be transferable to {\it concave} solute geometries
has been  anticipated by Carey, Chen,  and Rossky~\cite{Carey:2000} who concluded that "the enthalpy and entropy, associated with  displacement of the  solvent on substrate  binding [to a concave protein pocket] should not follow 'conventional' entropy driven hydrophobic  interactions."  Indeed,  experimental  studies of synthetic host-guest systems where hydrophobic confinement is
involved exhibit enthalpy-driven association~\cite{Smithrud:1991}. More recently, systematic molecular dynamics 
computer simulations by Setny, Baron, and McCammon have unequivocally demonstrated that the hydrophobic 
binding between generic uncharged ligands and pockets is indeed strongly driven by enthalpy and results
 from the expulsion of disordered water from the weakly hydrated cavity~\cite{Setny:2010}. The experimental 
relevance of this finding has been highlighted shortly after by Englert {\it {\it {\it et al.}}} who presented evidence 
that indeed the displacement of a few, unstructured water molecules from a hydrophobic pocket creates an enthalpic signature in the
binding of phosphonamidate to the thermolysin protein~\cite{Englert:2010}.  In fact, it becomes more and more evident that numerous apolar protein cavities may be weakly hydrated or even dehydrated~\cite{Abel2:2010} which has nontrivial effects on 
ligand binding affinity~\cite{Wang:2011}.  A better understanding of these unconventional processes is fundamentally important for the 
interpretation of  virtually all ligand-binding processes to hydrophobic pockets in molecular 
recognition~\cite{Gilson:2007,Mobley:2009,Hummer:2010}.  

In the theoretical description of apolar association and recognition, interface models based only on surface
tension and solvent-accessible surface area (SASA) have been traditionally surprisingly insightful, e.g., 
SASA-like models~\cite{Roux:1999,Gilson:2007,Mobley:2009} or scaled particle theory (SPT)~\cite{Ashbaugh:2001,Huang:2001,Ashbaugh:2006}.  However, they are expected to fail in cases where only a few, disordered water molecules are involved~\cite{Hummer:2010},  as in a highly concave, apolar pocket, and "the corresponding free energy seems unlikely to be suitably described 
by an  interfacial tension" \cite{Carey:2000}. To tackle this specific problem, Young {\it et al.}~\cite{Young:2007} introduced a 
method based on the displacement of quasilocalized waters upon ligand binding and recently  introduced a term attributable 
to the occupation of the dehydrated regions by ligand atoms~\cite{Wang:2011}. 

However, in this paper we argue that surface area based  models may remain valid even for high concave curvature if the antagonistic effects of  concave vs. convex bending on water interface thermodynamics are properly taken into account. 
Our reasoning is based on a generalization of capillarity theory, resembling
SPT, but extended to inhomogeneous and high local curvature and coupled to weak dispersion 
interactions~\cite{Boruvka:1977,Dzubiella:2006}.  Since this capillarity theory involves the minimization of
solute-solvent interfacial area, the existence of dehydrated ('dry') states of apolar pockets can be in principle 
captured~\cite{Setny:2009}. The basic model is briefly motivated in the next section while the thermodynamic
consequences and a detailed comparison to the computer simulations by Setny {\it et al.}~\cite{Setny:2010}  are discussed
in the Results section. The qualitative agreement between the interface model predictions and the simulations support
the general validity of our predictions down to subnanometer length scales. In particular, an important conclusion 
is that for fully complementary apolar pocket-ligand binding, water carries a similar thermodynamic signature as 
planar (large-scale) hydrophobic association  independent of the particular pocket/ligand geometry.

\section{Methods}

\subsection{Basic framework}

Let $\cal V$ be the volume occupied by water in three-dimensional space $\cal W$. 
In the volume void of water $\cal W\setminus\cal V$, the solvent is displaced by the solute atoms  and possibly by capillary effects. The  free energy of the  solvent interface surrounding the apolar solutes was proposed to be a geometric functional of $\cal V$ of the form~\cite{Dzubiella:2006}  
 \begin{equation}
G[{\cal V}] = P\int_{\cal W\setminus\cal V} dV +  \gamma\int_{\partial \cal V} [1-2\delta H(\vec r)] dA+\rho_0\int_{\cal V}U_{sw}(\vec r)dV
\end{equation}
where $\int_{\cal V}$  and $\int_{\partial \cal V}$ denote volume and surface integrals over $\cal V$ or its boundary $\partial \cal V$, respectively.
$P$ is the liquid bulk pressure, $\gamma(T)$ the temperature-dependent liquid-vapor surface tension,  $\rho_0(T)$ the water bulk density, and $\delta (T)$ the coefficient of the curvature correction to the surface tension linear in mean curvature $H(\vec r)$. The latter is defined by the mean of the two local principal surface curvatures $1/R_1$ and $1/R_2$, that is $H=(1/R_1+1/R_2)/2$.  In the last term in (1), the water density couples to the inhomogeneous, nonelectrostatic solute-water energy potential $U_{sw}(\vec r)$. Typically in molecular modeling, a Lennard-Jones (LJ) potential is used to express the Pauli repulsion and dispersion attraction between a solute atom and a water molecule. Thus, $U_{sw}(\vec r)$ may represent the sum of the LJ interactions of $N_s$ solute atoms fixed at positions $\vec r_i$ acting on bulk water possibly present at $\vec r$, i.e,  $U_{sw}(\vec r)=\sum_i^{N_s} U_{\rm LJ}(\vec r-\vec r_i)$.  Since the LJ coupling is proportional to density and  volume, it can be interpreted as a local mechanical pressure exerted by the solute atoms on the solvent interface. Without curvature correction and dispersion interactions, eq.~(1) reduces to the macroscopic interfacial free energy well known in thermodynamics since Gibbs~\cite{rowlinson}. 

\subsection{The curvature correction}
The curvature correction term linear in integrated mean curvature $H$ in (1) has been justified for various reasons. While  first derived by Tolman 1949 for spherical symmetry on thermodynamic grounds~\cite{Tolman:1949}, it is an integral part of the phenomenological SPT~\cite{Ashbaugh:2006} which has proven extremely successful in describing the solvation thermodynamics of spherical solutes in simple liquids.  SPT even serves as a good fit for the hydration free energy of  hard spheres in water~\cite{Lum:1999} and can be used for more sophisticated cavity hydration analysis using revised SPT~\cite{Ashbaugh:2006,Ashbaugh:2009}.  However, more than 30 years ago, Boruvka and Neumann extended Gibbs' classical theory of capillarity to arbitrary surface geometries having high, inhomogeneous curvature distributions~\cite{Boruvka:1977}. Within this generalization, the curvature correction in (1) was derived on the basis of fundamental equations of Gibbs' dividing surfaces making use of extensive geometric properties. In fact, a mathematical theorem well-known in differential geometry, the {\it Hadwiger theorem},  states that a functional defined on a set of three-dimensional bodies satisfying certain constraints, such as motion invariance, continuity, and additivity, must be a linear combination of the four mathematical measures volume, surface area, and the surface integral of mean and Gaussian curvatures~\cite{Hadwiger,Mecke}.  The Hadwiger perspective is actually equivalent to Boruvka and 
Neumann's generalized capillarity theory, and the three constraints above can be understood as a more precise definition for the 
conventional term {\it extensive}~\cite{Konig:2004}.

Indeed, the application of geometric functionals akin to (1) to {\it solvation processes} resulted into the first successful morphological  description of the thermodynamics of confined fluids, such as hard-sphere solvents~\cite{Konig:2004,Hansen-Goos:2007} or water in protein side chain packing~\cite{Yasuda:2010}.  In another line of developments, the  free energy (1),  {\it minimized} by $\delta G[{\cal V}]/\delta {\cal {V}}=0$ for a given solute geometry in water,  yielded a proper description of the  hydration free energy and interactions of small apolar solutes~\cite{Cheng:2007}. In particular, capillary evaporation between  extended hydrophobic plates found in computer simulations was quantitatively captured for the first time in an implicit description. Remarkably, the free energy and polymodal hydration behavior of water in the binding between an apolar pocket and a ligand found in MD simulations could be quantified and interpreted in terms of topologically distinct interfaces~\cite{Setny:2009}. In other words the (stable or metastable) 'dry' state observed in many protein cavities~\cite{Abel2:2010} is a local minimum of the interfacial free energy functional (1). These works underline the validity of the geometric functional and its potential not only to describe 'classical' macroscopic capillary effects but even for the quantitative or at  least qualitative description of aqueous hydration in small-scale apolar confinements of arbitrary geometry.

It is important to remark that the use of only the simple morphological measures linear in volume, surface area, and integrated curvature in (1) should be strictly valid  only if all intrinsic (correlation) length scales are small compared to the (confining) system size~\cite{Konig:2004}. Close to  the critical point, for instance, correlations become long-ranged and the geometric expansion in (1) may break down~\cite{Boruvka:1977,Stewart:2005}. Given the promising results cited above and presented in this work, however, the quantitative 
validity range of (1) for water at ambient conditions seemingly reaches to surprisingly small length scales.

\subsection{Obtaining thermodynamic coefficients by fitting to the hydration of LJ spheres}

In order to obtain an estimate for values of the correction coefficient $\delta(T)$
we minimize eq.~(1) and compare the solution to MD simulation results of the hydration free  energies of the 
relatively large (in terms of atomic sizes) apolar and spherical atomic solutes xenon and methane~\cite{Paschek:2004}. For spherical geometry
and negligible pressure contributions on small scales, eq.~(1) reduces to a function of the interface radius $R$ given by  
\begin{eqnarray}
G = 4\pi R^2\gamma(1-2\delta/R)+4\pi \rho_0\int_R^\infty U_{\rm LJ}(r)r^2dr.
\end{eqnarray}
The minimization $\partial G/\partial R=0$ yields
\begin{equation}
\frac{2\gamma}{R}\left(1-\frac{\delta}{R}\right)-\rho_0 U_{\rm LJ}(R) = 0 
\end{equation}
which is solved by the optimal interface radius $R_0$, and  $U_{\rm LJ}= 4\epsilon[(\sigma/r)^{12}-(\sigma/r)^6]$ is 
the Lennard-Jones (LJ) interaction. Plugging back $R_0$ into (2) gives the hydration free energy of the solute.  
The parameter $\delta(T)$ is then determined  by comparing the resulting hydration free energy to those  of methane and xenon obtained from explicit-water computer simulations for a variety of temperatures and water models~\cite{Paschek:2004}.  The input parameters in (2) are the liquid-vapor surface tension of a planar interface $\gamma(T)$  and the liquid density $\rho_0(T)$.  The values for $\gamma(T)$ and $\rho_0$ for SPC/E and TIP4P are taken from the works  of Chen and Smith~\cite{Chen:2007} and Paschek~\cite{Paschek:2004} and are summarized in Tab.~I.    This fitting procedure at $T=300$~K yields  $\delta = 0.086\pm0.009$~nm and $\delta = 0.076\pm0.008$~nm for SPC/E and TIP4P water, respectively, comparable to the values found in previous studies on the hydration of hard spheres~\cite{Huang:2001,Ashbaugh:2001,Floris:2005}.  From repeating this fitting procedure for temperatures $T=275$~K and $T=325$~K, we estimate the slope of $\delta(T)$ by the finite  difference $\partial \delta/\partial T (T=300~K) = [\delta(T=325~K)-\delta(T=275~K)]/(50~K)$ and find values  $T\partial \delta/\partial T = -0.090\pm0.003$~nm and $-0.144\pm0.005$ nm for SPC/E and TIP4P water at $T=300$~K,  respectively. All values found for $T=300$~K are summarized in Tab.~I. 

As a consistency check of the fitting procedure we have calculated the hydration entropy of OPLS methane~\cite{opls} using eq.~(4) below. Taking $R_0=0.31$~nm for OPLS methane from minimization of (2) and the TIP4P thermodynamic parameters at $T=300$~K, we obtain $TS=-14.2$~kJ/mol what indeed compares favorably with the result from Paschek~\cite{Paschek:2004}. Note that also the 
value of  $R_0$ is consistent with the effective radius estimated from water density distribution averaged in the MD simulations, see the discussion in III.C.  

\section{Results and discussion}

\subsection{Interface thermodynamics}

Let us now have a closer look on the thermodynamic consequences of description (1). Identifying (1) with a free energy in the 
Gibbs ensemble ($NPT$) with enthalpy and entropy  defined via $G=H-TS$, the entropy $S$ is 
given by~\cite{Boruvka:1977,Boruvka:1985} 
\begin{eqnarray}
S &=& -\left(\frac{\partial G}{\partial T}\right)_{N,P}\nonumber \\ &=&  -\frac{\partial \gamma}{\partial T} \int_{\partial \cal W} [1-2\delta H] dA+  2\gamma  \frac{\partial\delta}{\partial T} \int_{\partial \cal W}H dA, 
\end{eqnarray}
where the contribution of the van der Waals term in eq.~(1)  has been neglected.  At ambient conditions the relative water density changes with $T$ are typically very small, and it is well-known that van der Waals interactions contribute primarily to 
the enthalpy~\cite{Chandler:2005}.  Note that for the creation of a planar liquid-vapor interface the curvature terms can be neglected and simply $G=\gamma A$ and $TS =  -TA\partial \gamma/\partial T$. Given a surface with unit area $A=1$~nm$^2$ and the surface tension of real water at $T=300$~K, we  obtain $G = 43$~kJ/mol, $TS = 29$~kJ/mol, and $H = G+TS = 72$~kJ/mol.   Thus, we recall that the creation of a planar hydrophobic interface is strongly penalized by enthalpy which can be attributed to the breaking of hydrogen bonds. The water interfacial entropy increases due to 
the increased number of configurational degrees of freedom of the water molecules. This is well-known as the thermodynamic signature of large-scale hydrophobicity~\cite{Chandler:2005}.

What is the consequence of curvature? By definition the mean curvature $H$ changes sign when going from a convex surface to a concave one. Here, we follow the convention that a locally convex solute curvature (e.g., a spherical solute such as an argon atom) has a positive mean curvature, $H>0$. Consequently, $H<0$ for concave curvature (as for a droplet or a hemispherical protein pocket). With that fixed, the change of the free energy with curvature is governed by the correction coefficient $\delta(T)$. For real  water no experimental efforts have been undertaken so far to determine $\delta(T)$. However, good water models exists with respect  to the description of material and thermodynamics properties, such as the liquid density and the liquid-vapor surface tension. The SPC/E~\cite{berendsen:jpc} and TIP4P~\cite{tip4p} models are examples for such well-performing water models, and qualitative agreement to real water can be expected for curvature effects to the surface tension. By minimizing eq.~(1) and fitting it to hydration free energies of hydrophobic (hard-sphere-like or LJ) solutes obtained from explicit-water computer simulations (see Methods), we find {\it positive} values of $\delta=0.086$~nm for the SPC/E water model and $\delta=0.076$~nm for TIP4P at $T=300$~K. This is in good agreement with literature values where 0.09$\pm$0.02~nm are typically obtained by fitting (1)  to hydration free energies of hard spheres of a wide range of radii up to 1~nm~\cite{Huang:2001,Ashbaugh:2001,Floris:2005}.  As a first consequence,  the solvation free energy (1) per unit area  increases with increasing solute concavity and decreases with increasing solute convexity. Since $\delta$ is of \AA ngstrom size, the correction is notably for curvature radii below $\simeq 1-2$~nm, a realm where water interfaces are intrinsically rough due
to surface reconstructions of the hydrogen bonding network~\cite{felix:2009}. 

What about the balance between entropy and enthalpy? For this we need an estimate for the temperature-dependence of both the
surface tension $\gamma(T)$ and the coefficient $\delta(T)$, as emphasized by Ashbaugh and Pratt within SPT~\cite{Ashbaugh:2006,Ashbaugh:2009}. From the work of Chen and Smith~\cite{Chen:2007} we estimate  $T(\partial \gamma/\partial T) = -23.4$~kJ/mol/nm$^2$ for SPC/E water and $T(\partial \gamma/\partial T) = -36.7$~kJ/mol/nm$^2$ for TIP4P water for $T=300$~K. (For real water, $T(\partial \gamma/\partial T) = -27.9$~kJ/mol/nm$^2$).  From our fitting procedure (see Methods) for temperatures between 275 and 325~K, we find {\it negative} values for the slope of the coefficient $\delta(T)$ of about $\partial \delta/\partial T = - 4.8\cdot10^{-4}$ and $-3.1\cdot10^{-4}$ ~nm/K  for TIP4P and SPC/E at $T=300$~K, respectively. Ashbaugh and Pratt also found a negative slope of $\delta(T)$ of about $-7\cdot10^{-4}$~nm/K for hydration of a hard spherical cavity in SPC/E water. However, they employed a fitting formula different than (1) for large curvatures based on revised SPT~\cite{Ashbaugh:2006}.  Recently, Graziano used eq. (1) on Ashbaugh and Pratt's data to find  $\partial \delta/\partial T = - 3.7\cdot10^{-4}$~nm/K at $T=300$~K`\cite{Graziano}.

We first exemplify curvature effects for  spherical symmetry for which eq. (4) reduces to 
\begin{eqnarray}
S/A= -\partial \gamma/\partial T + (\partial \gamma/\partial T) 2\delta/R+  2\gamma (\partial\delta/\partial T)/R,
\end{eqnarray}
 where the radius $R>0$ for convexity and $R<0$ for concavity. Let us discuss numbers at hand of the popular SPC/E water model for which we obtain $TS/A= [23.4 - 4.0/R -6.8/R]$~kJ/mol/nm$^2$, where $R$ has the unit of 1~nm.  For convex solutes, where $R>0$, this implies the well-known change of sign of the solvation entropy when moving from large to small solutes at a crossover scale $R_c$.~\cite{Chandler:2005} We find $R_c\simeq 0.46$~nm for SPC/E water below which hydration thermodynamics carries the signature of small-scale hydrophobicity~\cite{Chandler:2005}, i.e., negative hydration entropies.  The existence of a crossover-scale larger than atomic radii may be special for water-like fluids as has been recently emphasized by Ashbaugh~\cite{Ashbaugh:2009} analyzing eq.~(5) taken from SPT. The special nature of the water hydrogen-bond network may lead to the relatively large (positive) value
 of $\delta$  when compared to LJ fluids~\cite{block,blokhuis} and contrasting the negative values of organic solvents~\cite{Ashbaugh:2009}.
 
However, for local solute concavity, per definition $R<0$, and the entropy per area (3) {\it increases} with $1/|R|$.  Because the total interfacial free energy $G/A$ rises with concave bending, increasing entropy also means a more strongly  increasing enthalpic penalty. In this perspective, the physical picture emerges that for concave bending of a planar liquid-solute interface more and more hydrogen-bonds adjacent to the bent interface are broken~\cite{Belch:1985}, thus the enthalpic solvation penalty grows, and more configurational freedom (disorder) adjacent to the interface is created. In other words,  friendly hydrogen-bonding neighbors are squeezed away in a concave environment and an interfacial water molecule has more freedom to fluctuate. The general trend makes sense as such as a extremely small apolar cavity would bear only one or two water molecules with a high entropy gain when compared to the more ordered bulk network. It indeed conforms with the experimental and simulation picture in which a few water molecules in strictly apolar cavities are typically characterized as being disordered. They possess enhanced configurational freedom (entropy) due to highly unsaturated hydrogen bonding as firstly shown by Chau for generic concave geometries~\cite{Chau:2001} followed by studies for more complex cavities~\cite{Carey:2000,Vaitheeswaran:2004,Setny:2006,Young:2007,Ewell:2008,Yu:2010,Setny:2010}. 

\subsection{Hydrophobic association}

The free energy of hydrophobic association, $\Delta G$, along a reaction distance $r$ between 
two approaching solutes is formally given  by the difference in hydration free energies between infinite separation 
and the final distance $r_0$, i.e.,   
\begin{eqnarray}
\Delta G &=& G(r=\infty)-G(r_0).  
\end{eqnarray}
Analogously it holds for the entropy $\Delta S = S(r_0)-S(r=\infty)$.  As frequently rationalized in literature large scale hydrophobic association is enthalpy driven, see e.g., the association of planar hydrophobic plates~\cite{Zangi:2008}, for which the curvature corrections in (4) hardly plays a role. In this case, $\Delta G \simeq \gamma\Delta A$, plus an enthalpic  van der Waals contribution, and $\Delta A$ is the desolvated plate area. As a consequence,  $\Delta S \simeq -\partial \gamma/\partial T \Delta A$ and we obtain the signature of large-scale association, that is enthalpy-driven.  Below a crossover scale, see the discussion above, the association of two small convex solutes is entropy-driven, as (convex) high-curvature parts of the solute-solvent interface are cut away upon association.
It is by now a well-established view in the physical and biological chemistry communities that small-scale hydrophobic association is always entropy driven.  That this view is generally not acceptable has been demonstrated  by computer simulations and experiments of small-scale apolar pocket-ligand binding~\cite{Smithrud:1991,Carey:2000,Setny:2010,Englert:2010}. 

To rationalize this within our perspective, let us first analyze the binding of a small solute (say methane) with radius $R=0.3$~nm 
and a {\it planar} hydrophobic interface. Let us assume that upon association half of the methane is desolvated and additionally 
a planar interfacial part of area $\pi R^2$. The water-contributed binding free energy would be $G_{ww} = -2\pi R^2 \gamma (1-2\delta/R) - \pi R^2\gamma$. Taking the values for SPC/E water at $T=300$~K we obtain $G_{ww} = -19.3$~kJ/mol. The entropy is given by $S_{ww}=2\pi R^2 [(\partial \gamma/\partial T)(1-2\delta/R)-2\gamma(\partial \delta/\partial T)/R] + \pi R^2(\partial \gamma/\partial T)$. Summing up, the result is a negligible $TS_{ww} = +0.6$~kJ/mol contributing to the total energy, i.e, the favorable entropic contribution from desolvation of the sphere and the unfavorable contribution from desolvation of the plane roughly compensate each other. Thus, the main driving force is enthalpy-driven. This in qualitative agreement with recent computer simulations on hydrophobic model enclosures, where apolar association turns from entropy-driven to enthalpy-driven when one of two small solutes is replaced by one or two planar hydrophobes~\cite{Abel:2010}. 

As argued above, concavity further increases the disorder of interfacal water with respect to a planar interface. Thus, 
in the geometric perspective formulated by (1), the puzzling thermodynamic controversy apparently  arising 
in small-scale apolar pocket-ligand binding can be resolved if the antagonistic 
effects of concave and convex bending on water interface thermodynamics are taken 
into account. If the (concave) pocket and the (convex) solute exactly complement each other geometry-wise, in other words, the surrounding solute-water interface  contours are identical, the curvature corrections in (1) and (4) cancel each other due to the opposite sign of the local mean curvature. Fig.~1 (a) and (b) show two illustrating sketches of spherical and a more complex pocket-ligand geometry, respectively. Consequently,  within perspective (1), {\it for fully complementary apolar pocket-ligand (pl) binding, water's thermodynamic signature is the same as for large-scale hydrophobicity}, i.e., 
\begin{eqnarray}
\Delta S_{\rm pl}/\Delta A  =  -\partial \gamma/\partial T, 
\end{eqnarray}
that is, dominated by enthalpy. $\Delta A$ is the total solute-solvent interface area which vanishes upon association, i.e.,
twice the desolvated area of the ligand-water interface upon fully complementary binding. 

Intriguingly, eq.~(4) predicts  that this result is generally  valid independent of the particular geometry of the pocket and ligand, as long they are fully complementary. However, this strong statement has to be weakened by following two notes.  First, as sketched in Fig. 1 (a) and (b) edge effects may occur resulting in nonvanishing corrections. However, these corrections can be included by a full numerical minimization of eq. (1) for a given geometry~\cite{Setny:2009}. Secondly, we emphasize again that (1)  should be strictly correct only for radii of curvature larger than typical correlation lengths in the fluid. Thus the quantitative validity of prediction (7) for small-scale pocket-ligand binding in water deserves further investigation.  

\subsection{Comparison to computer simulation results of apolar pocket-ligand binding}

In the following we attempt a more detailed comparison between the  geometric description (1) 
and the results from the simulation setup of Setny {\it et al.}~\cite{Setny:2010} . Local interface curvatures in the latter
are large and no quantitative agreement may be anticipated. However, the geometrical analysis will
be illustrative and will illuminate the qualitative validity of description (1) at hand of a well-defined 
model setup for apolar pocket-ligand binding.

To directly refer to the geometry found  in the MD simulation we estimate effective interface radii of the reference ($r=\infty$) and bound ($r=r_0$) states from average water  density distributions calculated in the MD simulation.  The density distributions by Setny {\it et al.}~\cite{Setny:2010} are reprinted in Fig.~2.  Spherical interface radii are then fitted to the loci $\vec r_{1/2}$ where the water density is half the bulk density  $\rho(\vec r_{1/2})\simeq\rho_0/2$.  The solute pocket features (hemi)spherical geometry and is larger than the ligand, see Fig.~2. From the water density distribution by Setny {\it et al.}~\cite{Setny:2010}, we 
estimate  an average concave radius of curvature of about $R_1 = - 0.7\pm 0.1$~nm for the pocket-water 
interface with the ligand being far away.  In the simulations  the ligand is represented  by a (convex)  sphere modeled by the OPLS united-atom  force field for  methane~\cite{opls} in TIP4P water.  From the average water density distribution one finds $R_2 = +0.30\pm0.01$~nm.  
 
 For the bound state, $r_0 \simeq-0.35$~nm (in Setny's definition $\xi \equiv r_0$). Upon ligand binding the concave pocket-solvent interface with radius $R_1$ is cut away by the ligand  with a chord of length $2l=1.08\pm0.04$~nm, as estimated from Fig. 2.  The replaced interface area is then given by  $2\pi |R_1| h$ with $h=|R_1|-\sqrt{R_1^2-l^2}=0.068$~nm.   Further inspection of Fig.~2 reveals that the ligand with interface area $4\pi R_2^2$ dehydrates while a new, nearly flat interface with area $\pi l^2$ is created.   
Application of (1) without the van der  Waals term and for spherical geometry together with (6) leads to the following expression for the binding free energy contributed by water-water (ww) interactions only:
\begin{eqnarray}
G_{ww} &=& U_{ww} - TS_{ww} \nonumber \\
&=& -2\pi |R_1| h \gamma (1-2\delta/R_1)-4\pi R_2^2 \gamma (1-2\delta/R_2)+\pi l^2\gamma, 
\end{eqnarray}
with $R_1<0$ and $R_2>0$. Analogously, the water-contributed binding entropy reads
\begin{eqnarray}
S_{ww} &=& 2\pi |R_1| h \left[\partial \gamma/\partial T (1-2\delta/R_1)-2\gamma(\partial\delta/\partial T)/R_1\right]  \nonumber \\ 
&+& 4\pi R_2^2 \left[\partial \gamma/\partial T (1-2\delta/R_2)-2\gamma(\partial\delta/\partial T)/R_2\right] \nonumber \\ 
&-& \pi l^2 \partial \gamma/\partial T. 
\end{eqnarray} 
Using the thermodynamic parameters summarized in Tab.~I for TIP4P at $T=300$~K, we find that the water-water contribution to the total free 
energy is $G_{ww}=-32.6\pm7.9$~kJ/mol, to the entropy $TS_{ww}=-16.6\pm12.8$~kJ/mol, and thus the water-water energy $U_{ww}=G_{ww}+TS_{ww}= -49.8\pm20.7$~kJ/mol. The error mainly reflects the uncertainty we assigned to the lengths $R_1$, $R_2$, and $l$. As a result, water's signature in the hydrophobic association of the pocket and the ligand is driven by enthalpy in qualitative agreement with the MD simulations.  In our perspective this is rationalized by the displacement of highly concave interface parts upon  ligand binding.  The actual numbers we found are not too far from those found by Setny {\it et al.} in their computer simulations,  where $U_{ww}=-37.0\pm17.3$~kJ/mol and $TS = -12.6\pm17.3$~kJ/mol.   Note, however, that the simple calculation above still draws an idealized geometry and additionally edge effects have been neglected. Thus, a serious comparison of numbers is a bit ambitious at that stage of modeling. As can be seen in Fig.~2,  for instance, the newly created interface is not totally flat but may exhibit some curvature which, however, is difficult to quantify given the statistical fluctuations in the density distributions from the MD simulations.

For a better quantitative performance check of description (1), a full minimization for the current pocket-ligand geometry must be performed.  This is feasible but requires more detailed numerical attention due to the appearance of two free energy branches of 
hydration for the unbound states~\cite{Setny:2006,Setny:2009}.  The two branches correspond to solvated (wet)  and desolvated (dry) states of the pocket separated by an energy barrier.  Additionally a thorough analysis of the van der Waals contribution to the enthalpy profile $H(r)$ would be  highly desirable.  While this is out of scope of the current communication, a few interesting conclusions on the performance of (1) may be drawn by employing  a few preliminary results~\cite{Setny:2009}: the pocket-water interface profiles resulting from the minimization of (1) for the unbound reference state ($r=\infty$) and the bound state ($r=r_0$) of Setny's pocket-ligand system  are also plotted  in Fig.~2.  We find that for the wet-pocket branch in the reference state, the radius of  curvature of the pocket-solute interface  is  $R_1=-0.55\pm 0.1$~nm.  For the dry-pocket branch in the reference state the interface is 
almost flat with  $R_1=-2.50\pm 0.2$~nm.  The MD simulation result consistently fits right between them as a result from 
sampling both branches in equilibrium~\cite{Setny:2006}.  For the bound state, minimization of (1) leads to a final, slightly concave interface with radius  $R=-1.5\pm0.1$~nm, see the bottom-right panel in Fig.~2. Compared to the MD results, the interface seems a bit shifted outwards from the pocket but with similar overall curvature. If we apply this curvature correction to the $\pi l^2$ term in (9) 
we obtain an additional $\simeq +4$~kJ/mol to $G_{ww}$ and $\simeq +10$~kJ/mol to $TS_{ww}$, still yielding a quantitative agreement within the MD error bar and not at all altering the qualitative picture.  Thus, the minimzation of (1) is fully consistent with the water density profiles and the thermodynamics calculated in the MD simulations.  The enthalpy-dominated thermodynamics can be rationalized by the antagonistic effects of the concave and convex association geometry. 

\section{Concluding remarks}

We have demonstrated that the remarkable enthalpy-driven hydrophobic association in apolar (convex) pocket - (concave) ligand
binding geometries can be rationalized by a surface-area model which properly accounts for the right sign and magnitude 
of the curvature correction contributions.  Only the fitting of $\delta(T)$ and its slope to hydration 
free energies of simple spherical solutes such as methane or xenon have been input to the model. The key prediction 
of this perspective is that water in fully complementary apolar pocket-ligand binding exhibits a thermodynamic signature very 
similar to large-scale (planar) hydrophobic association, that is, enthalpy-dominated, and hardly depends on the 
particular pocket/ligand geometry.  The trends discussed in this paper qualitatively agree with recent 
computer simulations~\cite{Abel:2010,Setny:2010} and thereby support the validity of our 
conclusions down to subnanometer length scales. 

One interesting remark concerns the validity of prediction (7) for solvents in general. As (1) is a totally generic, geometrical
description, the binding entropy in solvophobic pocket-ligand systems may be hardly geometry-dependent in any solvent, 
as long as the radii of the confining  curvatures  are larger than typical solvent correlation lengths. This can be easily tested with 
simple solvents, such as hard-sphere~\cite{roth:2008,keylock} or LJ solvents. For hard spheres the validity of the geometrical 
description (1) has been confirmed already for confining containers including concavity~\cite{Konig:2004}. For LJ solvents, 
where also capillary evaporation may occur~\cite{bolhuis}, the sign asymmetry in concave vs. convex solvation 
has unequivocally be proven by computer simulations~\cite{block} in agreement with microscopic statistical mechanics
approaches~\cite{blokhuis}. Note that the special nature of the water hydrogen-bond network may lead to the relatively large (positive) value
 of $\delta$  when compared to LJ fluids~\cite{block,blokhuis} and contrasting the negative values of organic solvents~\cite{Ashbaugh:2009}.

We have investigated purely apolar and mostly spherical geometries. The reality in 
biomolecular systems is considerably more complex. Protein pockets, for instance, ubiquitously feature local polar groups such as 
backbone carbonyls or weakly polar side chain groups. Their effects on water binding makes the interpretation
of the thermodynamic signature much more subtle due enhanced enthalpic contributions which in turn may
compensate entropy or not~\cite{Yin:2007,Young:2007,Qvist:2008,Yu:2010}.  Indeed, a single electrostatic charge placed in 
the simulated pocket or centered in the ligand of the system of Setny {\it {\it et al.}} may  significantly change water's entropic signature~\cite{Baron:2010}, even depending on the sign of charge. Additionally,   biological pockets exhibit strongly inhomogeneous high-curvature distributions. These subtleties obviously pose major challenges to surface-tension based hydration models as (1), 
in particular  with respect to the  inclusion of polarity~\cite{Dzubiella:2006b,boli}, higher order curvature corrections~\cite{Dzubiella:2006b}, or their numerical evaluation~\cite{Cheng:2007}.  Our work also suggests that traditional SASA models~\cite{Roux:1999}, which are often applied
to predict binding free energies in molecular recognition processes~\cite{Gilson:2007,Mobley:2009}, may require revision of  
surface definitions with respect to local curvature for improving their quantitative performance and predicting binding
entropies. 
\\\\
{\bf Acknowledgements}
J.D. thanks the Deutsche Forschungsgemeinschaft (DFG) for support within the Emmy-Noether
Program and Piotr Setny, Riccardo Baron, and Felix Sedlmeier for helpful comments. 
J.D. is particularly grateful to Zhongming Wang for sending the pocket-ligand interface 
figures~\cite{Setny:2009} and to Piotr Setny and Ric Baron for sending the high-resolution
figure of water density distributions~\cite{Setny:2010}.

\newpage 

\begin{table}[h]
\begin{center}
\begin{tabular}{|c|c|c|c|c|c|}
\hline
 & $\rho_0$ & $\gamma$ & $T\partial\gamma/\partial T$ & $\delta$ 
& $T\partial \delta/\partial T$  \vspace{-0.2cm} \\ 
 & [nm$^{-3}]$ & [kJ/mol/nm$^2$] & [kJ/mol/nm$^2$] & [nm]  & [nm] \\
\hline
SPC/E  &  33.3  & 36.8 & -23.4 & 0.086 & -0.093 \\
\hline 
TIP4P  &   33.1  & 32.8 & -36.7 & 0.076 & -0.144 \\
\hline
\end{tabular}
\end{center}
\caption{Water parameters used in this work for a temperature $T=300$~K for the SPC/E and TIP4P models. 
The water liquid bulk density $\rho_0$ is taken from the work of Paschek~\cite{Paschek:2004}. The liquid-vapor surface tension $\gamma(T)$ 
and its slope have been taken from Chen and Smith~\cite{Chen:2007}. The curvature parameter $\delta(T)$ and its slope have been calculated by 
minimizing and fitting eq.~(2) to the hydration free energy of xenon and methane for temperatures 275, 300, and 325~K obtained
from explicit-water computer simulations~\cite{Paschek:2004}.}
\label{table2}
\end{table} 

\newpage 

\begin{figure}[htp]
\caption{Illustrative sketches of fully complementary apolar pocket-ligand association geometries. Convex interface 
curvature are shown by lines colored in green while concave interface parts are colored in red. In (a) the
ligand and pocket are both spherical with the same absolute radius $|R_1|$ and fully complement each other
upon binding, apart from small edge effects. In (b) pocket and ligand geometries are more complex but cancel
each other upon binding, so that the final state is indistinguishable from (a). In both cases the intrinsic entropic
signature of the water is predicted to be given mostly by  $\Delta S_{\rm pl}/\Delta A  =  -\partial \gamma/\partial T 
$, independent of the particular ligand geometry and scaling with the desolvated interface area $\Delta A$. }
\end{figure}

\begin{figure}[htp]
\caption{Color-coded water density ($\rho^*$) distributions from MD computer simulations of the generic pocket-ligand system by Setny {\it et al.}~\cite{Setny:2010}. The pocket-ligand distance $\xi$ along the $z$-axis is equivalent to reaction-coordinate $r$ in the main text. 
The spherical ligand with interface radius $R_2>0$ (thick green line) desolvates upon binding (top left to bottom right). At the same time it displaces the concave hemispherical pocket-water interface with radius $R_1<0$ (thick black line), with $|R_2|<|R_1|$, leaving an almost flat interface (dashed black line in the bottom right panel) with length $2l$. The interface profiles drawn by red lines are the solute-solvent interface lines obtained from a full minimization of eq.~(1) for the wet and dry pocket reference states (top left) and the bound state (bottom right)~\cite{Setny:2009}.  }
\end{figure}

\newpage
\begin{figure}[htp]
\includegraphics[width=12cm,angle=0]{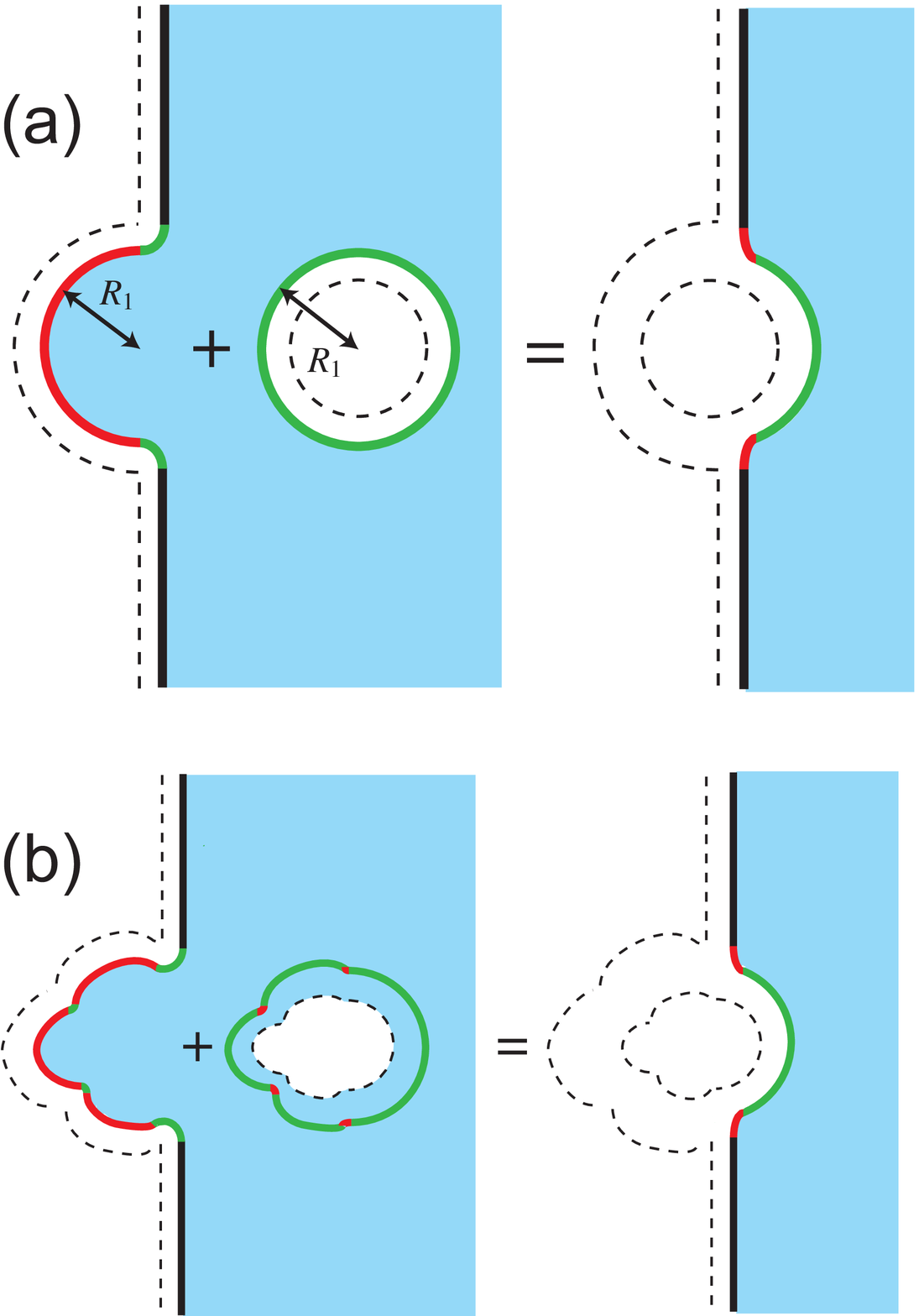}
\end{figure}

\newpage
\begin{figure}[htp]
\includegraphics[width=15cm,angle=0]{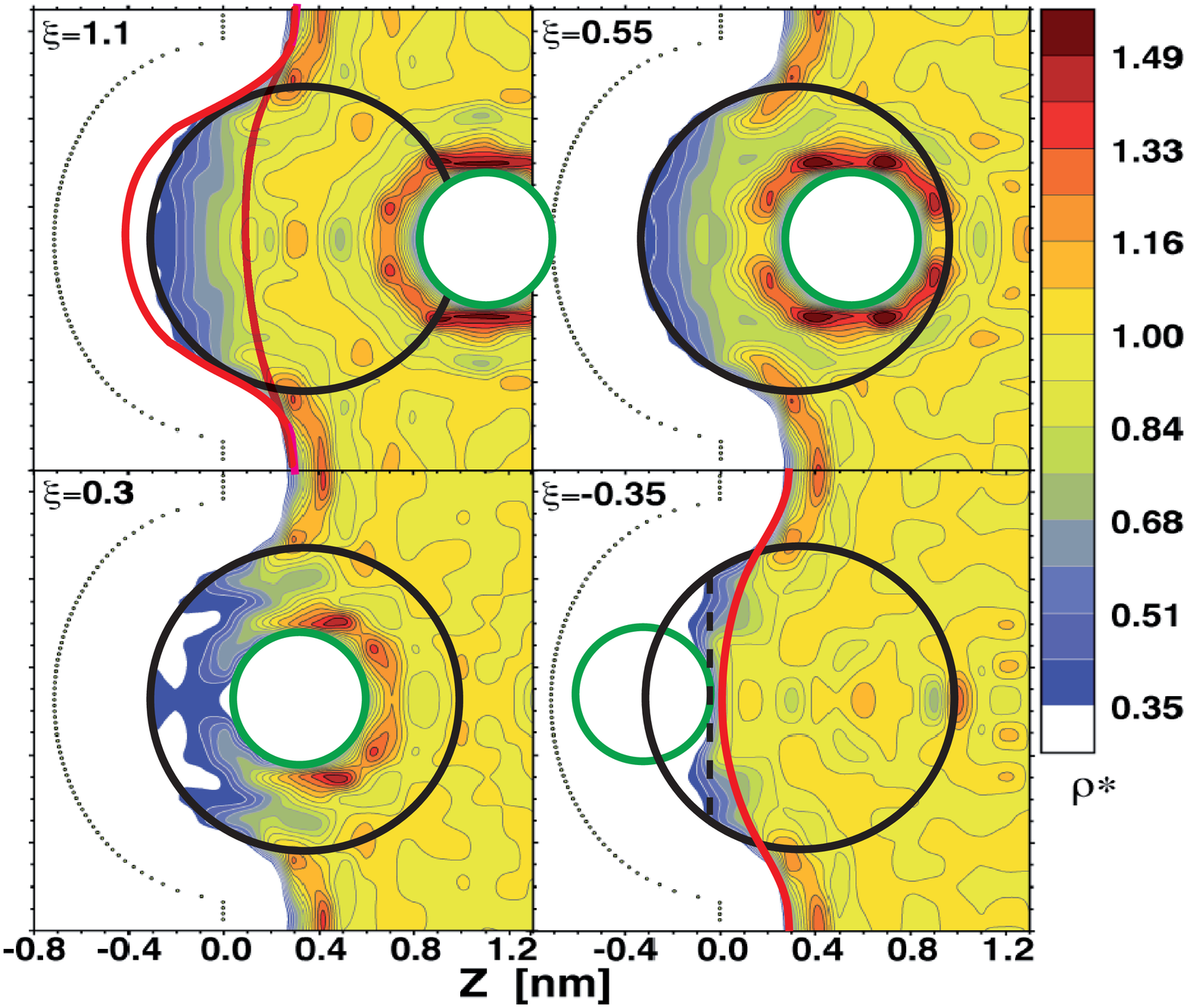}
\end{figure}

\end{document}